# Hacktive Matter: data-driven discovery through hackathon-based cross-disciplinary coding

Megan T. Valentine[1] and Rae M. Robertson-Anderson[2]

[1] Department of Mechanical Engineering, University of California, Santa Barbara, Santa Barbara CA 93106

[2] Department of Physics and Biophysics, University of San Diego, San Diego, CA 92110

Correspondence

valentine@engineering.ucsb.edu

randerson@sandiego.edu

## Abstract

The past decade has seen unprecedented growth in active matter and autonomous biomaterials research, yielding diverse classes of materials capable of flowing, contracting, bundling, de-mixing, and coalescing. These innovations promise revolutionary applications such as self-healing infrastructure, dynamic prosthetics, and self-sensing tissue implants. However, inconsistencies in metrics, definitions, and analysis algorithms across research groups, as well as the high-dimensionality of experimental data streams, has hindered the identification of performance intersections among such dynamic systems. Progress in this arena demands multi-disciplinary team approaches to discovery with scaffolded training and cross-pollination of ideas, and requires new methods for learning and collaboration. To address this challenge, we have developed a hackathon platform to train future scientists and engineers in 'big data', interdisciplinary collaboration, and community coding; and to design and beta-test high-throughput (HTP) biomaterials analysis software and workflows. We enforce a flat hierarchy, pairing participants ranging from high school students to faculty with varied experiences and skills to collectively contribute to data acquisition and processing, ideation, coding, testing and dissemination. With clearly-defined goals and deliverables, participants achieve success through a series of tutorials, small group coding sessions, facilitated breakouts, and large group report-outs and discussions. These modules facilitate efficient iterative algorithm development and optimization; strengthen community and collaboration skills; and establish teams, benchmarks, and community standards for continued productive work. Our hackathons provide a powerful model for the soft matter community to educate and train students and collaborators in cutting edge data-driven analysis, which is critical for future innovation in complex materials research.

1. **Introduction**

Active materials can do amazing things: change shape and size, generate mechanical forces, and sense and respond to external stimuli, with potential to revolutionize applications to self-healing infrastructure, soft robotics, and biosensing. One driver of the unprecedented advances in the design, synthesis and characterization of soft active materials is the ongoing innovation in computation and data science, which has unlocked new abilities in high throughput (HTP) data acquisition and analysis, and in turn, enabled data-driven approaches to modeling. Together these advances provide insight into increasingly complex materials formulated across large compositional and formulation spaces. Within the US, the Materials Genome Initiative (MGI) provides an organizing framework for researchers, educators and funding agencies to develop the infrastructure and tools needed to enable and accelerate data-intensive approaches to materials discovery[1]. Although MGI-informed studies have increased data accessibility and the use of data-driven modeling, we lack formal education in data science within most physics and materials communities, and integrating MGI training into research-driven projects remains challenging, limiting forward progress[2].

The properties of soft active materials themselves offer unique challenges and opportunities due to their out-of-equilibrium responses and structural and dynamical heterogeneity[3-9]. Structural arrangements on molecular to mesoscopic scales occur over a wide range of timescales, arising from intrinsic biomolecular kinetics or responses to external signals, generating incredibly complex patterns and functions that enable, e.g., crawling, morphogenesis, and phase separation. The primary datasets used to characterize such materials are typically microscopy videos that are each 10s of GBs in size. The vast number of components and possible interactions in these systems makes it essentially impossible for a single team to fully explore the accessible parameter space, which would also require dozens of TBs of data. Yet, collaboration is challenging since the size and complexity of data limits the ability to easily share original files between groups, and makes processing images in a manageable time frame difficult, often leading to important information being ignored or discarded. Moreover, there is little consensus in metrics, definitions, and analysis algorithms across research groups, further hindering the identification of performance intersections among materials.

To address these needs, we designed a multiyear hackathon platform to enhance skill and professional development within our collaborative team (**Fig 1**), while generating accessible tools to enable: (1) sharing and analyzing large, expansive, and disparate datasets, and (2) establishing a common language and metrics for describing soft active materials. We focused on active cytoskeletal composites (ACCs), which are in vitro networks of actin filaments and microtubules, two principal biopolymers comprising the cell cytoskeleton, that are acted on by their respective molecular motor proteins, myosin and kinesin, to undergo contraction, flow and restructuring[3, 4, 6, 8, 9]. We aimed to uncover intersections between different data sets and methods by arming trainees with approaches to efficiently screen, process and categorize big data. We chose to implement hackathons as scalable, low-risk training platforms to teach researchers how to analyze complex and heterogenous data to enable decision making while providing them with experience and skills in team coding, user interfaces, and best practices for algorithm development. By having all participants work together to create a single collaborative end product, we aimed to instill a strong sense of belonging and accountability, while providing a shared understanding among our collaboration of how our research fit into the larger MGI community and goals. Our hackathon framework, consisting of tutorials, small group coding sessions, breakouts, and large group discussions, not only advanced software development, but strengthened collaboration skills and team interactions as well as trainee interest and skills in active soft matter research. Below, we describe our approaches, results and lessons learned to enable broader use of hackathon platforms to promote training and innovation in complex materials science.

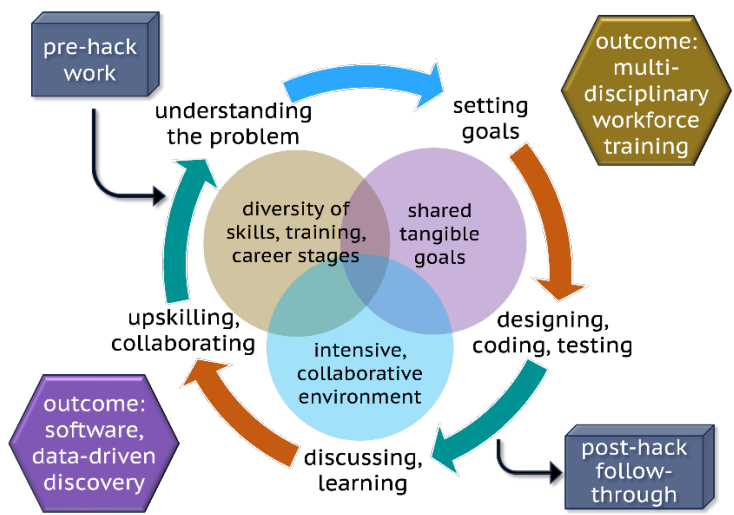

Figure 1. Goals, workflow and outcomes of hackathons for training and discovery in interdisciplinary active materials research. The center translucent circles indicate the essential components of the hackathon, and the circling arrows indicate the workflow. Blue boxes indicate the importance of pre- and post- activities, and hexagons indicate key training and discovery outcomes.

## 2. Hackathons as a model for innovation and collaboration

Hackathons are time-bounded collaboration events that bring together disparate groups to work intensely toward shared goals[10-12]. The platform originated within the open-source software community, to broaden access and thereby accelerate code improvements. More recently, hackathon events have evolved into a more general platform for innovation and interdisciplinary science[12-19]. A key feature is their use of "radical collocation", in which team members work in close proximity, and nearly continuously during the project duration, which has been shown to boost productivity for time-bounded work[20]. Participants in hackathons have described a unique 'feeling' and 'culture' of such events, which, unlike traditional workshops or conferences, are focused on collectively generating solutions rather than transferring existing knowledge[21-23]. Hackathons have been shown to stimulate creative thinking and accelerate discovery, while providing immediacy and excitement to research outcomes[10]. They also boost morale through a shared sense of short-term accomplishment, and enable problem solving for urgent problems[16].

A defining feature of hackathons is that participants are provided with an opportunity to work collectively and in a new environment on tasks that they are passionate about, free from usual constraints and interruptions[11]. By producing an open, respectful, and fun environment in which all participants are expected to contribute, hackathons flatten the playing field[19, 23]. This structure encourages diverse experts to interact outside of their specific domain(s) and allows them to develop solutions and uncover pitfalls with minimal risk[11]. Engaging the whole group minimizes the risk of 'free-riders' and promotes transparency in work allocation and deliverables[24, 25]. Hackathons are also an excellent training tool, allowing rapid education in areas of emerging knowledge and urgent need[15, 19]. The classical classroom is completely reframed as an interactive, team-driven social event, with self-directed learning, broad sharing of knowledge, and collaboration at all stages of work[26, 27]. Teachers become facilitators, who can adapt the learning environment to meet student needs on demand. In this role, educators often ask questions and innovate methods outside of their own core training, which is critical to advancing solutions in rapidly developing fields[10, 19, 28].

Hackathons also expose students to design reasoning paradigms and allow them to identify and solve problems in a human-centered context[27, 29]. While concepts of team building, problem identification, brainstorming, ideation, prototyping, and production are commonly taught in engineering disciplines, such skills are often underdeveloped in physics and materials-centered curricula. The flexible and adaptive nature of hackathons enables iterative refinement as the understanding of both the problem and solution coevolve[29]. Such approaches have been reported to provide value throughout project lifetime, enabling the development of skills and pipelines in early stages that are later leveraged to create community-centered tools and methods that are eventually deployed to solve pressing problems[14].

In order to maximize the benefits of hackathons, it is important to identify and addresses common challenges. Hackathons are resource intensive in terms of physical infrastructure, hardware and software access, and the need for highly engaged facilitators so site selection and planning are essential [21, 24]. The short project horizons and focus on solutions can limit design choices and hinder deep thinking about the problem[27, 29]. Moreover, the flat hierarchy and open access model boosts creativity, but may lead to gaps in required subject matter expertise, and the time-limited team interactions may leave some questions unanswered by the end of the event[10, 14, 19, 29]. We demonstrate that through intentional planning, including pre- and post-event activities, these risks can be mitigated, making hackathons an extremely useful tool for active materials research.

### 3. Guiding research needs for developing a hackathon platform

Our motivation was the need for our multi-institution, multidisciplinary collaboration to develop a unified workflow for screening and analyzing large volumes of data to enable material optimization toward specific performance metrics. Our primary data were fluorescence microscopy videos of complex cytoskeletal active matter, with multiple components (e.g., biopolymers, crosslinking proteins and molecular motors) labeled with distinct fluorophores. Our trainees ranged from high school students to postdocs with backgrounds in physics, engineering, biophysics and statistics; some with no prior coding experience and others who were very skilled, although in different coding languages (MATLAB, Python, C++). Given the diversity and complexity of material responses that we can access using active cytoskeleton composites (ACCs), we desired a consistent framework for characterization and analysis that could be easily implemented across our five separate physical locations, and that everyone could understand and use.

Our initial attempts at coordination involved having researchers from each group try to measure a specific metric at their own campus, for example the speed of contraction in an actively remodeling ACC, which they would then report and compare. Although trainees were encouraged to collaborate and share codes, the outcomes were inconsistent and difficult to interpret. Different trainees and labs used different approaches ranging from particle-image velocimetry (PIV), to differential dynamic microscopy (DDM), to particle-tracking, all of which require a number of subjective choices and inputs with no unifying guidelines[3, 30, 31]. Digging deeper, we found gaps in the foundational understanding of the underlying algorithms and assumptions, which made it difficult for teams to understand what methods were best suited for what types of analyses and why. Moreover, even within our relatively small collaboration, we lacked a common language to discuss our approaches to analysis and material performance. For example, terms like 'error', 'noise' and 'uncertainty' have many meanings when applied to large and complex microscopy videos. Our team consisted of physicists, engineers, statisticians, and biologists all of whom had different ideas, tolerances, and approaches for quantifying and handling such effects.

We faced similar challenges in describing the material performance. In general, the quantification of mechanical properties from video-based data is challenging, relying on models and assumptions that our typical active systems violated. Seemingly simple tasks, such as determining if a material was sufficiently 'active' and 'resilient' to be prioritized for follow-up analyses, were challenging. While we generally agreed that 'active' materials displayed athermal directed motion, we lacked workflows and criteria for deciding if and when a material exhibited and maintained activity, or robust definitions of what made one formulation more active than another. In defining 'resilience', we struggled to reconcile the concepts of temporal resilience of long-lived percolated networks with elastic resilience of energy absorbing structures. It was clear from these initial attempts that we needed clearer definitions, unifying workflows, and common analysis platforms to enable material design, optimization and analysis across our team. To address these needs and kickstart our collaborative work on unifying our approaches to complex material analysis, we launched a multiyear effort resulting in a robust hackathon platform to enhance training and collaboration within our team.

## 4. Hackathon Design

**4.1** *Logistics.* Hackathons are exciting and effective tools that emphasize creativity, flexibility, and communication to enable complex multidisciplinary problems to be tackled. However, the need to quickly pivot to meet unexpected challenges and needs requires a different approach to planning and execution than a typical conference or workshop. The selection of the physical site is important[12, 32]. Each year we gathered in person at the University of California, Santa Barbara in Elings Hall, home of the California NanoSystems Institute, for two to three days during summer when students and faculty had more flexibility in their schedules and the campus was less crowded, to enable more time and space to focus on our collective goals. The venue enabled all sessions to be held in the same building, and the location of the Institute provided access to walking trails and beachside open spaces for easy exploration during breaks. One large lecture-style meeting room served as the central hub where all large group discussions were held, with multiple breakout rooms and informal meeting spaces and lounges also available for the entirety of the hackathon for small group hacking sessions and tutorials (**Fig 2**). Each space had adequate electrical outlets and access to high-speed internet with sufficient bandwidth to enable work with HTP datasets and codes, and each room was Zoom-enabled, with ample seating and whiteboard space. While the meeting was intended to be in-person, each year we had some virtual participants, and we also used the Zoom capabilities to allow participants to easily share their screens during large group report-outs and to facilitate communication among groups located in different parts of the building. The large open workspace enabled each team to engage with other teams, and allowed facilitators to easily interact with all participants and each other and quickly communicate information and updates to the entire group[21, 33]. Our collaboration already had a suite of remote collaboration tools we regularly used, including Slack channels, GitHub repositories, and shared cloud-based Google Drives that we heavily leveraged to enable real-time communication and file sharing during the hackathon. These platforms were also critical for documenting code, capturing questions and problems, sharing successes and documenting critical information that we could retain, assess and revisit from year to year[23]. Because of the nature of our academic collaboration, we did not need to address any issues surrounding IP generation or use of proprietary data, which is a limiter for some hackathon designs[23].

| time | Year 1 Day 1 | Year 1 Day 2 | Year 2 Day 1 | Year 2 Day 2 | Year 3 Day 0 | Year 3 Day 1 | Year 3 Day 2 |
|---|---|---|---|---|---|---|---|
| 8am | Breakfast | Breakfast | Introductions, Overview, Goals | Day 1 Recap, Day 2 Goals | | Breakfast, Tutorial 2: Hands-on training with HTP software | Day 1 Recap, Day 2 Goals |
|  | Introductions, Overview, Goals | Day 1 Recap, Day 2 Goals | Participant Pre-work Presentations | Hack 3 | | | Tutorial 4: Database Design |
| 9 | Lecture 1 | Guided Group Work 3 | | | | Tutorial 3: working with common | Report Outs |
|  | Discussion 1 | | | | | | Tutorial 5: Simulations |
|  | Break | Break | Break | Break + informal discussions | | Break | |
| 10 | Guided Group Work 1 | Break | Guided Group Work 1 | Guided Group Work 3 | | Guided Group Work 1 | Break |
|  | | Hack 2 | | | | | Hack 3 |
| 11 | Lecture 2 | Report Outs | | | | | |
|  | Discussion 2 | | Report Outs | Report Outs | | Report Outs | |
| 12 | Lunch | Lunch | Lunch | Lunch | Travel, Check-in | Lunch | Lunch, Report Outs |
| 1pm | Lecture 3 | Guided Group Work 4 | Hack 1 | Hack 4 | | Hack 1 | Hack 4 |
|  | Discussion 3 | | | | | | |
| 2 | Guided Group Work 2 | Break | | | | | |
|  | Break | Hack 3 | Break | Break | | Break | Break, Report Outs, Discussion |
| 3 | Hack 1 | | Guided Group Work 2 | Hack 5 | | Guided Group Work 2 | Guided Group Work 3 |
| 4 | | | | | | | |
|  | Report Outs, Day 2 Planning | Report outs, Next Steps, Survey | Report Outs | Report outs, Next Steps, Survey | | Report Outs | Report Outs, Next Steps, Survey |
| 5 | Walk + Dinner + Informal Discussions | Off-site Dinner | Walk + Dinner | Off-site Dinner | Dinner, Introductions, Overview, Goals | Dinner + Networking + Brainstorming | Off-site Dinner |
| 6 | | | | | | Hack 2 | |
| 7 | | | Hack 2 | | Tutorial 1: Intro to HTP screening | | |
|  | | | | | Discussion, Q&A | | |
| 8 | | | Report Outs, Day 2 planning | | | Day 2 planning | |

**Figure 2. Summary schedules for annual hackathons in Years 1-3.**

As is typical for hackathons, copious amounts of coffee, snacks, and food were available throughout the meeting, and the importance of communal and accessible sustenance is reflected in participant responses to the post-survey question about what their favorite aspect of the hackathon was. Most meals were provided on campus to foster collaboration and community-building[23]. In Years 1-2 we provided meals at set times and chose outdoor venues a short walk from the meeting room to provide a new environment as well as fresh air and exercise. In Year 3, we aimed to provide more flexibility, and had food delivered to our headquarters at meal times so participants could take a break and eat when it worked best for them. We still encouraged taking meal breaks, eating outside, taking walks, etc. but there were not set times that everyone was forced to adhere to. In all years, the final dinner was off-campus and was a celebration of everyone's hard work. Specific schedules for each hackathon are summarized in **Figure 2**.

**4.2 *Laying the foundation*.** In year one, we focused on (1) understanding how, when and why the various algorithms and approaches work to enable informed decisions about their use and results; and (2) converging on a common language for discussing the properties of the active systems and the computational approaches to quantifying them. To narrow the scope to something we thought could be reasonably achieved in a 2-day event, we focused on using DDM to analyze data, and provided pre-hackathon reading assignments to build some common understanding among the group.

We took a scaffolded approach to the hackathon design, interweaving technical tutorial-style talks to lay the foundation and situate everyone on a level playing field, breakout groups on related topics, and small group coding challenges that first leveraged existing code and then focused on writing new code to analyze data (**Fig 2**). To provide trainees with a broader and more encompassing understanding of DDM and its uses and applications, we invited several faculty, postdocs and graduate students from outside of the collaboration, who were DDM experts, to present tutorials, serve as breakout group leads and participate in coding challenges. Each session was scheduled for 45 – 90 mins depending on the topic and goal, and the presentations and hands-on challenges started with the basics and built up to open-ended questions.

For the 'hacking' sessions – during which participants collectively wrote and tested software – we split the trainees into groups of 3-4 and tasked each with a design challenge to address a specific problem, which we had cooperatively identified in the large group discussions on the first morning. We designed groups to ensure a variety of coding expertise and skills in each, such that everyone could meaningfully contribute: from data curation and processing, to coding, to beta-testing and documenting. Each session had 2-3 assigned faculty leads who floated between the different groups to provide guidance, gauge progress and problems, and engage in discussions. After each discussion or hacking session we organized a report-out, where each participant or group was asked to provide a short update of their learning or results, and facilitators asked follow-up questions, noted synergies and connected groups to additional resources as needed. At the end of the hackathon, we set goals for integrating the methods learned into our research workflows, but did not define ongoing project-based work at this stage.

**4.3 *Coding it up*.** Having developed common language and understanding among participants, and preliminary frameworks and coding algorithms to assess material performance, we designed the Year 2 hackathon to be even more hands-on, with fewer tutorials and methodology discussions, and more emphasis on developing a uniform code for material screening (**Fig 2**). We also began to address the issues of efficient processing, screening, analysis, and sharing 'big data'. In an effort to create tools that would be broadly accessible to the greater materials science community we aimed to develop algorithms that were material-agnostic and did not rely on knowing anything about the physics or formulation of the system. While the first hackathon focused on using DDM to analyze data, here we expanded our toolkit to include higher throughput methods and those that assess a broader palette of properties. At this stage, we narrowed the participant list to 10 researchers, ranging from high school students to postdocs, who had substantial prior coding experience; and we included more substantial pre-hackathon work and a virtual preparation meeting.

Roughly one month before the hackathon, we provided participants with a curated set of 10 videos that were representative of the types of material systems we were investigating, and exhibited a variety of dynamical and structural characteristics. We established the coding goal of developing screening algorithms to enable various formulations to be rapidly assessed and optimized according to four performance metrics: contraction, stiffness, resilience, and long-range stress propagation. We challenged each participant to identify parameters and approaches to assess and quantify the given performance metrics in a way that limits subjectivity and streamlines screening. Specifically, we asked them to choose any four of the ten videos to analyze and then answer a series of questions, via an online form, including describing the video, suggesting ways to screen those videos worthy of further analysis, characterizing the data using one or more of the performance metrics, and deciding what results to save. We asked each participant to prepare 3 slides describing their approach and results, and submit them before the hackathon.

On the first morning of the hackathon we presented an overview of the goals of our collaborative research and the immediate aims of the hackathon to orient the participants to our shared goals. We then had a series of short presentations during which all participants presented the results of their pre-hackathon work. Based on these kick-off presentations, and the skillsets and interests of the participants, we formed three groups of 3-4 participants to tackle the three most promising of the originally-identified performance metrics (eliminating long-range stress transmission).

The remaining schedule was more flexible than in Year 1, with no predetermined tutorials or topical discussions. We instead aimed to foster collaboration, creativity and productivity by allowing coding tasks and goals to emerge more organically, initially guided by the participant presentations, then through interspersed large group discussions and report-outs and PI facilitators floating between small groups. We scheduled 8 hacks that were each ~90 mins and separated by ~15-30 min breaks and report-outs. We also added an evening session to the first day to make better use of the short time frame and acknowledge that different people are more or less productive at different times of day (**Fig. 2**). Some groups worked very collaboratively during the hacks while others preferred more independent work, splitting up tasks and reconvening at regular intervals to discuss progress. The format and space accommodated both approaches. If we identified a need for a short tutorial, the facilitators formed a subgroup, led by faculty or more experienced participants, and the group worked through the details at white boards until the issue was resolved. There were times when participants hit a roadblock that they couldn't overcome in the allotted time, in which case the faculty facilitators worked to find other ways for them to contribute, e.g., beta-testing, literature review, note-taking, data curation.

At the conclusion of the hackathon, the three groups presented their results and their planned next steps, which included validating their codes and making them more user friendly. We collectively established a series of benchmarks to achieve in the following weeks, including improving the throughput and integrating the three nascent software packages into a single framework, requiring collaboration among the different groups. Over the next year, slow but steady progress was made through a series of virtual meetings and continued collaboration. Software originally written in a variety of languages (Python, MATLAB, C++) were converted to Python, which the group agreed would be the most accessible to the community. We also submitted an abstract for two participants to present a poster about the code at the APS March Meeting the following spring, providing a fixed date deliverable that spurred progress[34]. The presentation also provided valuable feedback from the community.

**4.4 *Beta-testing, validating, and training.*** Building on the successes of the first two hackathons and the continued progress made in the months following, in Year 3 we were in a position to beta-test the high throughput screening algorithms we had developed to identify bugs and inconsistencies, test that outputs were consistent with qualitative observations, optimize efficiency, and ensure applicability for a wide range of active matter video data. We hoped to expand beyond the initial 3 metrics that we focused on in Year 2, and wanted to engage more trainees with different skill levels and coding experience. As in Year 1, we invited all members of our collaboration, regardless of their coding background.

To facilitate meaningful engagement of participants with varied backgrounds, we identified four participants to act as facilitators and help lead the pre-hackathon work. We focused on packaging the existing codes into a single executable with reasonable and understandable inputs and outputs and an interface that users of all coding levels could execute. We curated a large (~200 GB) shared dataset, with sufficient diversity and complexity of structures and dynamics, for all participants to work on to ensure consistency and enable comparison. We instructed each participant to download the Python software packages to their individual laptops prior to the hackathon, and on the first day provided them with a thumb drive that contained the curated dataset.

Based on survey feedback from the previous years, we built more structure into the schedule than we had in Year 2, with more defined short term goals, but continued to focus on hands-on hacking rather than formal presentations (**Fig. 2**). We added a working dinner on the day that participants arrived, to allow everyone to get to know each other and to provide an overview of the hackathon goals and how they relate to our collaboration goals. We confirmed that everyone had downloaded the data and codes they needed, and then the two trainees who led the development of the unified software package presented a basic tutorial of how to operate the code. This evening introduction session allowed us to dive right into a hands-on training session the following morning, where all participants attempted to run the analysis code on a common small test set of videos, with close guidance from the developers. This exercise was extremely helpful not only to ensure everyone was on the same level playing field to use and develop the software, but it also revealed a number of bugs that cropped up only on certain Python distributions and/or operating systems, which the developers were able to immediately address in real-time. We then had the trainee who led data curation describe the rich dataset that all the participants would be working on, explain the material system and common types of observable behavior, and suggest some exemplary videos to use for initial testing. The rest of the meeting was a series of 90-minute hacking sessions and 30-minute report-outs and group discussions (**Fig 2**). Rather than defining goals for each hack, we empowered participants to define their own schedules and benchmarks based on our end goal of producing a user-ready HTP software that we could use amongst our collaboration and share with others in the field.

The software we were testing comprised three complementary branches that leveraged distinct image analysis algorithms, so we divided trainees into three groups to beta-test, validate, debug, and enhance each branch. Each team had a mix of coding expertise from novices to those who helped develop the code. Over the course of the two days, what began with a software that screened for 3 parameters using binary yes/no outputs, evolved into software that reported more than ten continuous parameters that were output in a table and a graphical display. In the several months following the hackathon, a subset of participants worked to refine and optimize the code, BARCODE: Biomaterial Activity Readouts to Categorize, Optimize, Design and Engineer, which is now publicly available to all those interested in the high throughput screening and characterization of dynamically restructuring soft materials[35-37].

## 5. Outcomes

**5.1 *Products:*** The primary aim of our hackathons was to develop unified workflows, algorithms, and language for processing and analyzing video-based data of active matter systems, with an eye towards HTP screening for specific performance targets. After three consecutive annual hackathons, we ultimately produced a functional software package: BARCODE[36, 37]. This software package integrates the creative ideas and custom code of many trainees, and could not have been produced in a cohesive, vetted way without the dedicated time and space that a hackathon allows. Hackathon activities and results from each year also independently led to publications, conference presentations, new insights, and accelerated research progress that fed into the final product.

**5.2** *Collaboration and Upskilling:* We primarily assessed program success and learning outcomes through internal evaluation, using anonymous surveys we conducted at the end of each hackathon. These results were used to guide the design of subsequent hackathons. We aimed to increase collaboration and meaningful interactions between trainees from different institutions, at different levels and with different backgrounds, and to improve data literacy and computational analysis skills to prepare trainees for the new workforce. Each hackathon formed new connections and strengthened existing ones between trainees, as well as the facilitators, which enabled continued collaborations following the hackathon. The survey results, summarized in **Figure 3**, corroborated the success of these goals. Respondents consistently reported increases in understanding, interest, skills and confidence as a result of their participation, nearly half of the respondents anticipated using what they learned immediately upon their return to their labs, and >80% anticipated using what they learned within the following year.

A key feature of our hackathons is the shared goal of all participants, who collaborate rather than compete, ensuring everyone has a vested interest and can meaningfully contribute. Participants reported that they enjoyed these aspects of the work, including learning how others solved the problems, getting feedback on their work, and orienting their efforts to align with the common goals (**Table 1**).

| | |
|---|---|
| I liked the goal setting…as well as reporting in about my progress and hearing about what other people had worked on. I also enjoyed learning about how people from other groups were approaching problems. | I really enjoyed the collaborative aspect of the workshop. It was great to talk to the other researchers and hear about their work. It was also interesting hearing what problems people were facing and the resulting group discussions to help figure out that problem. |
| Group discussions and post-lecture discussions were great. enjoyed hearing many different perspectives on understanding the same systems. I think every individual's expertise was taken advantage of in a very productive way and it remained an interesting and engaging experience for everyone. | I enjoyed meeting other members of the group and improving collegiality. I appreciated the opportunity to readily exchange information and ideas in person. I felt this interaction allowed me to progress much faster than I otherwise would have in understanding the different research projects in the collaboration, and how I can contribute. |
| I really just liked all the feedback, and all our open discussions. I feel like that is where I learned the most when we were brainstorming, and thinking through why things would work and why other methods would not. | The collaborative nature…kept it engaging, and enabled me to make greater progress than I feel like I would have accomplished when working on my own, or even in a smaller group. |
| The codes improved a lot. I made use of most of time efficiently without [it] feeling boring. | Learning and improvement of computational skills |
| Having time to brainstorm new approaches | Learning from others regarding how they analyze the problem. |

**Table 1**: Selection of responses to the following prompt: "Please share some of your favorite aspects of the workshop"

**5.3** *Educational Outcomes*: The open-ended problem-centered nature of hackathons fosters critical thinking, teamwork and communication, with discussion-based learning happening within and between teams[14, 38]. Participants learn to be nimble – working through multiple cycles of problem identification, skill building, and execution – while learning how to pivot or adapt when faced with roadblocks[38]. These phases of brainstorming, ideation, and invention are essential skills for innovators[24]. Participants also developed experience working in skill-diverse teams, and learning to manage and resolve intergroup conflicts[21, 39]. Participants gained practical hands-on training in image-based data analysis, HTP screening, and managing big data, aligned to training the MGI workforce. We anticipate that the hackathon experience will also empower participants to provide unique perspectives on future product design and development[29].

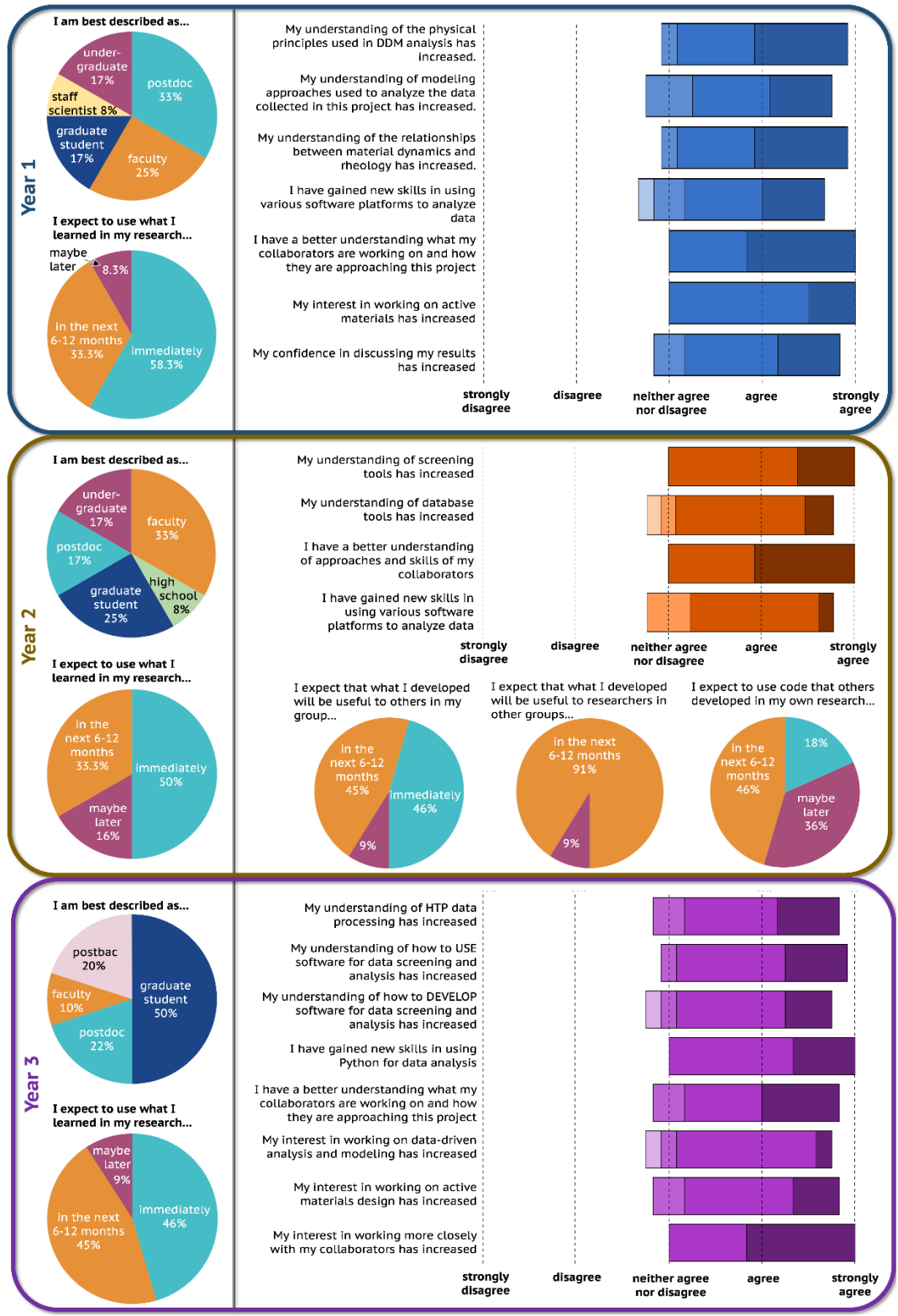

**Figure 3:** Summary of program evaluation data. Number of respondents was 12, 13, and 12 in Years 1, 2, and 3, respectively.

## 6. Insights

When we first envisioned running a series of hackathons to help our collaboration develop skills and code for active material analysis, we were not sure how they would impact the researchers in our groups, and whether we would be able to meaningfully advance our research and training goals in such a short time period. What we found was that hackathons could be designed to be extremely valuable tools for collaborative science and engineering, and that useful products and positive learning outcomes could result. Here, we summarize some of the insights we developed through our design and execution, some of which we discussed above, to enable others to incorporate hackathon-based learning and projects into their own research.

**6.1** *Pre-hackathon planning:* We aimed to create an environment and activities that would spur collaboration and creativity and promote transferrable skills, while advancing common coding goals. To this end, facilitators typically started meeting 6 months prior to the event to begin to plan the technical and learning goals, develop pre-hackathon activities for participants, and handle various logistics. Prior to each hackathon, we assigned homework to the participants related to the goals of that event, and provided links to relevant online resources including Python tutorials and example codes. This type of pre-event engagement has been shown to improve hackathon success[19]. In Year 1, we focused mainly on technical readings and practice codes related to the tutorials presented. In Year 2, we crowdsourced the brainstorming and ideation phases by asking participants to begin to independently solve problems of video processing using curated datasets, which in turn informed our on-site down-selection and accelerated the prototyping of solutions that the whole group would pursue. By Year 3 we had developed specialists among the participants, who were becoming experts in the approach and execution, and they were tasked with leading the final stages of prototyping in advance of the hackathon so that our on-site group work could focus on debugging and validating our product.

**6.2** *Setting the tone:* We aimed to create an environment that valued risk-taking and framed failure as a normal part of the hacking process rather than a sign of inferior competence or skill[10, 11]. In our kick-off sessions, we discussed our goals of collaboration, rather than competition, and explained that this was a supportive space to try out new ideas, make mistakes, and learn from them[10, 24, 33, 40]. We purposefully invited participants from a wide range of educational backgrounds and skill levels, creating teams that balanced beginners and experts to enhance idea exchange and creativity[12]. We provided beginners with additional training and tutorials, and encouraged experts to take leadership roles in explaining concepts and writing more complex code[11]. Certain essential concepts, for example command line scripting or using a Github interface, were presented to the whole group, with hands-on examples that each individual executed on their own laptop. We encouraged discussion and training exchanges within the group but did not move on to the next activity until we confirmed that everyone was able to complete the basic task, ensuring that everyone had the basic skills needed to fully participate[28].

We considered the skill diversity within our group of participants to be a strength – providing avenues for out-of-the-box thinking and solutions that a group of experts may not have considered. In order to ensure that all participants could meaningfully contribute, we aimed to select problems with enough complexity and dimensionality that multiple solutions of varying degrees of technical difficulty were possible[21]. An advantage of our hackathons was that nearly all participants had been members of our collaboration for some time, so everyone had a shared knowledge of the active material system, the general structural and dynamic properties the networks displayed, and the overall goals of the research, providing an important common grounding among the group.

**6.3 *Timing considerations:*** We chose to schedule our hackathons in the summer when both students and faculty have more flexibility in their schedules, and campus rooms are more accessible. We found that ~2.5 days was optimal to achieve our goals within logistical and budgetary constraints, and ensure participants were productive and engaged the entire time. Shorter times would not have allowed us to lay the groundwork and then build up to the final projects, with ample time for thinking, discussions, debate, brainstorming, and reporting. By the end of the second full day, most participants were demonstrating a decline in creativity and stamina that suggested that further work would likely lead to diminishing returns. We wanted participants to leave feeling 'exhausted in a good way'–having learned a lot, developed new skills, advanced their research, and solidified new and existing collaborations–but also energized and inspired to continue working on the problems we tackled during the hackathon, both independently and in collaboration with other trainees across our multi-institution collaboration.

Prior to the hackathon, we provided participants with a structured schedule; however, we were intentionally flexible in our execution to allow new topics or needs to be introduced and addressed, and to make adjustments on the fly to foster productivity. The faculty facilitators, who were actively engaged with the participants throughout the event, played a key role. We defined expectations and goals at the kickoff session of the meeting that served as a North Star for decision making during the hackathon, and we worked together to identify cases where flexibility or adaptation was required. Changes were often driven by participants, and decided collaboratively in our large group report-outs. They typically took the form of continuing projects from the previous 'hack' because certain participants and/or groups were 'in the zone' and making progress. We did not want to slow their momentum and force them to switch gears in order to start on the originally planned new problem. When a change was needed, we provided rationale and advanced notice so participants could plan accordingly and continue to focus on the most important immediate goals[23]. By Year 3, we introduced a lot of flexibility into the schedule, including flexible break and meal times so that participants could choose their own time to shift gears or take a break. While this change worked well for some, allowing them to retain focus and momentum at critical points, others reported finding it hard to take a break if not mandated or if they saw others working.

Overall, the 'living' nature of the event is a hallmark of hackathons: the rapidly changing format provides motivation and excitement, and accelerates problem solving[24]. It also, however, challenges facilitators, who need to be able to communicate, shift gears, and adapt with little advanced planning and often based on ideas or needs that emerge organically from the participants. We also learned the importance of careful balance between structured and open-ended work. Participants enjoyed the extended time to work on problems, but preferred having clear goals for each session. Given the different skill levels, some participants needed help identifying very specific tasks, while others enjoyed the freedom to come up with solutions that address broader goals, as reflected in participant feedback on program design (**Table 2**).

| |
|---|
| I really liked the fact that we had so much time to just try to implement and code it up. |
| We have a clear goal to work on and to finish coding sessions. |
| I thought the focus was suitably narrow to allow ideation and pursuit of an attainable objective. |
| Working on a sub part of the screening was nice since that let us dig deep into the code |
| I appreciated the ability to test the code on diverse use cases to test code design and screening heuristics. |
| Forming groups of 2-4 and giving plenty of time to work with share-outs every ~2 hours seemed to really work well! |
| The diversity of experiences was helpful in getting an understanding of the goals of the group. |
| I appreciate you gave me introduction papers to read before…to comprehend more about the lectures. |

**Table 2**: A sampling of participant feedback on program design

**6.4 *Continuity:*** Even the most successful hackathons generally fall short of delivering an optimized product ready to deploy, so we intentionally planned for continuation of projects started during the hackathon in the months to follow. We ended each hackathon with presentations by participants who demonstrated the code, workflows, and/or analysis that they had developed, described what was still needed, and outlined their plans to complete or build on their results. Some presentations were highly collaborative, with groups presenting a single framework and plan, while others were more independent, with each group member presenting on a different aspect of the problem they were tackling. We found it important to allow for and even encourage this flexibility due to different working styles and skillsets of participants. During these presentations, we had extended discussions as a large group to give feedback, help refine plans, and find synergies and potential redundancies between different approaches. We ended each event with a concrete plan and goal for each participant or group to accomplish over the following 2-3 months. We had several virtual meetings following each hackathon to encourage and guide continued work and collaboration.

The multi-year collaborative nature of our research certainly helped in sustaining our efforts, as we had a dedicated group of PIs committed to working together on research questions in active matter, and we met regularly as a group to advance this work. Despite relatively high trainee turnover from year to year, we found a surprising level of continuity in skills and understanding. We attribute this to the outreach of former participants who upon return to their home campuses became ambassadors of the project, transferring their knowledge within our larger collaboration, thereby helping to build institutional knowledge and maintain momentum, even among team members who did not participate in a given hackathon. To onboard new participants each year, we took care in designing the pre-hackathon work to set expectations and orient all participants to the problems we aimed to tackle, and provided tutorials and resources explaining the approaches that could be used to solve them. This allowed newcomers to rapidly get up to speed, while bringing fresh perspectives and enthusiasm that enriched the learning environment and outcomes.

The biggest post-hackathon push occurred after Year 3, led by participants who had developed enough expertise to lead the post-hackathon work of finalizing the software package and disseminating the results. It is common for the post-hackathon work to be more successful when the product is more developed and there is clear consensus that the added effort will lead to successful project completion, including the eventual deployment of the software or technology[41]. This can lead to a tradeoff between hackathons focused on individual skill development, and those focused on producing a working prototype[42]. We found our multi-year, tiered approach, rooted in our collaborative research, provided the right balance of learning and production while allowing our team to meet our growing research goals.

**7. Future opportunities:** We plan to continue using hackathons as a model for training scientists and engineers in data science and collaborative research. In future years we have goals of incorporating more modeling and simulations into the toolbox of approaches, and incorporating AI and ML methods to categorize and cluster data and inform the experimental design. We also plan to host hackathons beyond our immediate collaboration to increase cross-pollination of ideas and standardize frameworks and best practices for active matter characterization. Finally, we hope to expand this teaching approach to develop non-technical skills such as writing papers or proposals and preparing presentations. We believe similar design principles could be highly effective in training in these skills, with a common goal defined at the onset, pre-event preparation and post-event follow-up work. These professional development focused hackathons could include an intensive bootcamp of short tutorials, bursts of writing and/or preparing presentations, peer reviews, collaboration and large discussions where participants share their results. Overall, the hackathon model developed here provides a powerful tool for the soft matter community to train researchers in collaboration, coding, and data-driven analysis, to drive the frontiers of innovation in soft complex materials research.

**Acknowledgments**

We acknowledge funding from the US National Science Foundation DMREF program through following grants, including MGI supplemental funding: NSF DMR-2119663 (to RMRA), NSF DMR-2118497 (to MTV). We acknowledge the NSF BioPACIFIC Materials Innovation Platform (DMR-1933487) for providing data science infrastructure and training, and for use of the microrheology suite. We thank Christopher Dunham, BioPACIFIC MIP, for guidance in database development; Ryan McGorty, Moumita Das, and Mengyang Gu for their contributions to planning and facilitating the Hackathon events; the California NanoSystems Institute at UC Santa Barbara for logistical support; and all Hackathon participants for their creative contributions to this work.

**References**

1. *Materials Genome Initiative for Global Competitiveness*, US National Science and Technology Council, 2011.
2. J. J. de Pablo, N. E. Jackson, M. A. Webb, L.-Q. Chen, J. E. Moore, D. Morgan, R. Jacobs, T. Pollock, D. G. Schlom, E. S. Toberer, J. Analytis, I. Dabo, D. M. DeLongchamp, G. A. Fiete, G. M. Grason, G. Hautier, Y. Mo, K. Rajan, E. J. Reed, E. Rodriguez, V. Stevanovic, J. Suntivich, K. Thornton and J.-C. Zhao, *npj Computational Materials*, 2019, **5**, 41.
3. R. J. McGorty, C. J. Currie, J. Michel, M. Sasanpour, C. Gunter, K. A. Lindsay, M. J. Rust, P. Katira, M. Das, J. L. Ross and R. M. Robertson-Anderson, *PNAS Nexus*, 2023, **2**.
4. G. Lee, G. Leech, P. Lwin, J. Michel, C. Currie, M. J. Rust, J. L. Ross, R. J. McGorty, M. Das and R. M. Robertson-Anderson, *Soft Matter*, 2021, **17**, 10765-10776.
5. D. A. Fletcher and P. L. Geissler, *Annual Review of Physical Chemistry*, 2009, **60**, 469-486.
6. T. E. Bate, M. E. Varney, E. H. Taylor, J. H. Dickie, C.-C. Chueh, M. M. Norton and K.-T. Wu, *Nature Communications*, 2022, **13**, 6573.
7. S. Banerjee, M. L. Gardel and U. S. Schwarz, *Annual Review of Condensed Matter Physics*, 2020, **11**, 421-439.
8. J. Alvarado, M. Sheinman, A. Sharma, F. C. MacKintosh and G. H. Koenderink, *Nature Physics*, 2013, **9**, 591-597.
9. J. Sheung, C. Gunter, K. Matic, M. Sasanpour, J. L. Ross, P. Katira, M. T. Valentine and R. M. Robertson-Anderson, *Macromolecular Rapid Communications*, **n/a**, 2401128.
10. J. Falk, A. Nolte, D. Huppenkothen, M. Weinzierl, K. Gama, D. Spikol, E. Tollerud, N. P. C. Hong, I. Knäpper and L. B. Hayden, *IEEE Access*, 2024, **12**, 133406-133425.
11. E. P. P. Pe-Than, A. Nolte, A. Filippova, C. Bird, S. Scallen and J. D. Herbsleb, *Journal*, 2019, **36**, 15-22.
12. D. Huppenkothen, A. Arendt, D. W. Hogg, K. Ram, J. T. VanderPlas and A. Rokem, *Proceedings of the National Academy of Sciences*, 2018, **115**, 8872-8877.
13. A. Richterich, *Convergence*, 2019, **25**, 1000-1026.
14. A. Ghouila, G. H. Siwo, J.-B. D. Entfellner, S. Panji, K. A. Button-Simons, S. Z. Davis, F. M. Fadlelmola, T. D. o. M. H. Participants, M. T. Ferdig and N. Mulder, *Genome Research*, 2018, **28**, 759-765.

15. A. L. Ferguson, T. Mueller, S. Rajasekaran and B. J. Reich, *Molecular Systems Design & Engineering*, 2019, **4**, 462-468.
16. T. Feder, *Physics Today*, 2021, **74**, 23-25.
17. G. Mulholland and B. Meredig, *MRS Bulletin*, 2015, **40**, 366-370.
18. K. M. Jablonka, Q. Ai, A. Al-Feghali, S. Badhwar, J. D. Bocarsly, A. M. Bran, S. Bringuier, L. C. Brinson, K. Choudhary, D. Circi, S. Cox, W. A. de Jong, M. L. Evans, N. Gastellu, J. Genzling, M. V. Gil, A. K. Gupta, Z. Hong, A. Imran, S. Kruschwitz, A. Labarre, J. Lála, T. Liu, S. Ma, S. Majumdar, G. W. Merz, N. Moitessier, E. Moubarak, B. Mouriño, B. Pelkie, M. Pieler, M. C. Ramos, B. Ranković, S. G. Rodriques, J. N. Sanders, P. Schwaller, M. Schwarting, J. Shi, B. Smit, B. E. Smith, J. Van Herck, C. Völker, L. Ward, S. Warren, B. Weiser, S. Zhang, X. Zhang, G. A. Zia, A. Scourtas, K. J. Schmidt, I. Foster, A. D. White and B. Blaiszik, *Digital Discovery*, 2023, **2**, 1233-1250.
19. A. Happonen, D. Minashkina, A. Nolte and M. A. M. Angarita, *AIP Conference Proceedings*, 2020, **2233**.
20. E. H. Trainer, A. Kalyanasundaram, C. Chaihirunkarn and J. D. Herbsleb, presented in part at the Proceedings of the 19th ACM Conference on Computer-Supported Cooperative Work & Social Computing, San Francisco, California, USA, 2016.
21. C. Wallwey, M. M. Longmeier, D. Hayde, J. Armstrong, R. Kajfez and R. Pelan, *Frontiers in Education*, 2022, **7**.
22. M. Rys, *Convergence*, 2022, **28**, 1800-1825.
23. L. Garcia, E. Antezana, A. Garcia, E. Bolton, R. Jimenez, P. Prins, J. M. Banda and T. Katayama, *PLOS Computational Biology*, 2020, **16**, e1007808.
24. M. Rys, *Knowledge Management Research & Practice*, 2023, **21**, 499-511.
25. J. Lobbe, B. Florence and J.-C. and Sagot, *International Journal of Design Creativity and Innovation*, 2021, **9**, 119-137.
26. S. Mhlongo, K. E. Oyetade and T. Zuva, 2020.
27. N. M. Martins Pacheco, M. Geisler, M. Bajramovic, G. Fu, A. Vazhapilli Sureshbabu, M. Mörtl and M. Zimmermann, *Design Science*, 2024, **10**, e9.
28. J. Wyngaard, H. Lynch, J. Nabrzyski, A. Pope and S. Jha, 2017.
29. M. Flus and A. Hurst, *Design Science*, 2021, **7**, e4.
30. R. Cerbino, F. Giavazzi and M. E. Helgeson, *Journal of Polymer Science*, 2022, **60**, 1079-1089.
31. J. C. Crocker and D. G. Grier, *Journal of colloid and interface science*, 1996, **179**, 298-310.
32. L. G. Morales, N. H. A. Savelkoul, Z. Robaey, N. J. Claassens, R. H. J. Staals and R. W. Smith, *PLOS Computational Biology*, 2022, **18**, e1009916.
33. M. Juraschek, L. Büth, N. Martin, S. Pulst, S. Thiede and C. Herrmann, *Procedia Manufacturing*, 2020, **45**, 43-48.
34. R. Robertson-Anderson, J. Michel, Q. Chen, K. R. Peddireddy, M. Rust, J. Ross, M. Das, R. McGorty and M. Valentine, *Bulletin of the American Physical Society*, 2024.
35. livingbam.org, 2025.
36. github.com/softmatterdb/barcode, 2025.
37. Q. Chen, A. Sriram, A. Das, K. Matic, M. Hendija, K. Tonry, J. L. Ross, M. Das, R. J. McGorty and R. M. Robertson-Anderson, *arXiv preprint arXiv:2501.18822*, 2025.
38. C. La Place, S. S. Jordan, M. Lande and S. Weiner, Columbus, OH, 2017.
39. M. Lara and K. Lockwood, *TechTrends*, 2016, **60**, 486-495.
40. M. Schneider, *Inc. Magazine*, 2017.
41. A. Nolte, I.-A. Chounta and J. D. Herbsleb, *Proc. ACM Hum.-Comput. Interact.*, 2020, **4**, Article 145.
42. A. Nolte, E. P. P. Pe-Than, A. Filippova, C. Bird, S. Scallen and J. D. Herbsleb, *Proc. ACM Hum.-Comput. Interact.*, 2018, **2**, Article 129.